\documentclass[reprint,
amsmath,
amssymb,
aps,prl,preprint,
groupedaddress,
onecolumn]{revtex4-2}
\usepackage{graphicx}
\usepackage{dcolumn}
\usepackage{bm}
\usepackage{subfigure}

\begin{document}

\preprint{APS/123-QED}
\title{Vanishing of Dimits Shift in Realistic Fusion Plasmas with Negative Magnetic Shear}


\author{Dingkun Yang $^{1}$}
\email{21936039@zju.edu.cn}
\author{Shengming Li$^{1}$}
\author{Yong Xiao$^{1}$ }
\email{Corresponding author:\\
yxiao@zju.edu.cn}
\author{Zhihong Lin$^{2}$}

\affiliation{\small $^{1}$Institute for Fusion Theory and Simulation, Department of Physics, Zhejiang University, Hangzhou 310027, China. \\
$^{2}$ Physics and Astronomy, UC Irvine, CA 92697, USA.}

\date{\today}

\begin{abstract}
This study employs gyrokinetic simulations to investigate ion temperature gradient (ITG) turbulence in realistic fusion plasmas featuring reverse magnetic shear. Negative magnetic shear is found to suppress the ITG instability due to the scarcity of mode rational surfaces, as evidenced by a comparison of instabilities for different magnetic shears. This suppression effect remains observable in nonlinear turbulence with zonal flow artificially eliminated, where the emergence of turbulence solitons aligns with mode rational surface peaks. However, the suppression effect diminishes in the presence of self-consistently generated zonal flow, along with the occurance of turbulence solitons.  The zonal flow is found to originated from a force driven process by the primary instability, instead of the conventional modulational instability. The study further reveals a remarkable phenomenon that the Dimits shift no longer exists for negative magnetic shear, which are attributed to the weakness of zonal flow around marginal stability. However, away from marginal stability, the turbulent transport is primarily regulated by the zonal flow regardless of different magnetic shears. 
\end{abstract}

\maketitle



Turbulent transport driven by instabilities in drift wave range is widely observed in experimental and space plasmas\cite{horton1999drift}. One of the common causes of ion turbulent transport in tokamaks is the Ion Temperature Gradient (ITG) instability \cite{coppi1967instabilities,romanelli1989ion,dong1992toroidal,dorland1993gyrofluid}, which significantly reduces the efficiency of fusion reactors. The profile stiffness propels fusion plasmas toward marginal stability \cite{doyle2007plasma}, resulting in the plasma edge temperature on the pedestal top being determined by the critical temperature gradient, as is the core plasma temperature. Consequently, even a slight elevation of the critical temperature could yield a remarkable increase in fusion power.

The axisymmetric $E \times B$ poloidal rotation or zonal flow, self-consistently generated by turbulence, is generally accepted as a regulator of the saturation level of turbulence and transport in tokamaks \cite{lin1998turbulent,rosenbluth1998poloidal}. The generation mechanism of zonal flow is usually attributed to the Reynolds stress via modulational instability \cite{chen2000excitation} or Kelvin-Helmholtz (KH) instability \cite{rogers2000generation}.

Near the marginal stability, the zonal flow has been observed to completely suppress turbulence and transport, causing a nonlinear up-shift of the critical temperature gradient for ITG, known as the Dimits shift\cite{dimits1996scalings,dimits2000comparisons}. Despite its significance in understanding crucial fusion scenarios, such as zonal flow-turbulence interaction and L-H transition \cite{diamond2005topical}, the physics mechanisms underlying the Dimits shift remain largely elusive. Conventionally, the Dimits shift has been attributed to tertiary instability  by the gyrofluid theory—a KH-like instability driven by strong zonal flow shear \cite{rogers2000generation}. However, a recent study based on the slab geometry and Hasegawa-Wakatani model proposes an alternative tertiary instability that shares the same instability drive as the primary instability \cite{zhu2020theory}. Notably, the Dimits shift is found to exist not only in collisionless plasmas but also in collisional plasmas \cite{mikkelsen2008dimits}, despite the collisional damping of the zonal flow\cite{hinton1999dynamics, xiao2007effects}.

\begin{figure}
     \centering
      \subfigure{
            \includegraphics[width=0.45\linewidth]{
    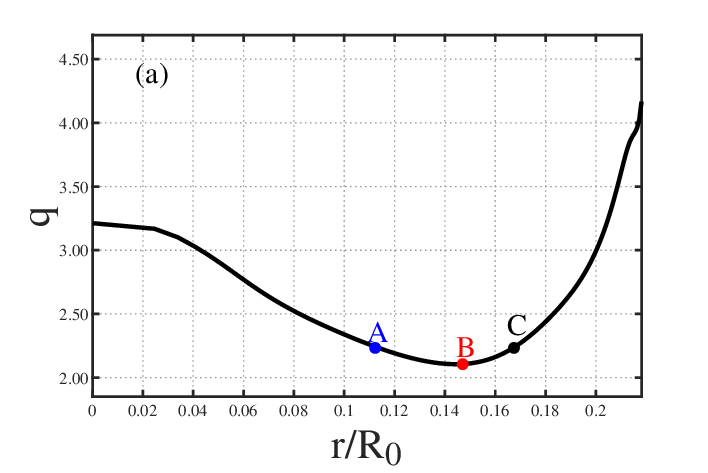}}
 \hspace{0.21in}
      \subfigure{
            \includegraphics[width=5cm,height=8cm]{
      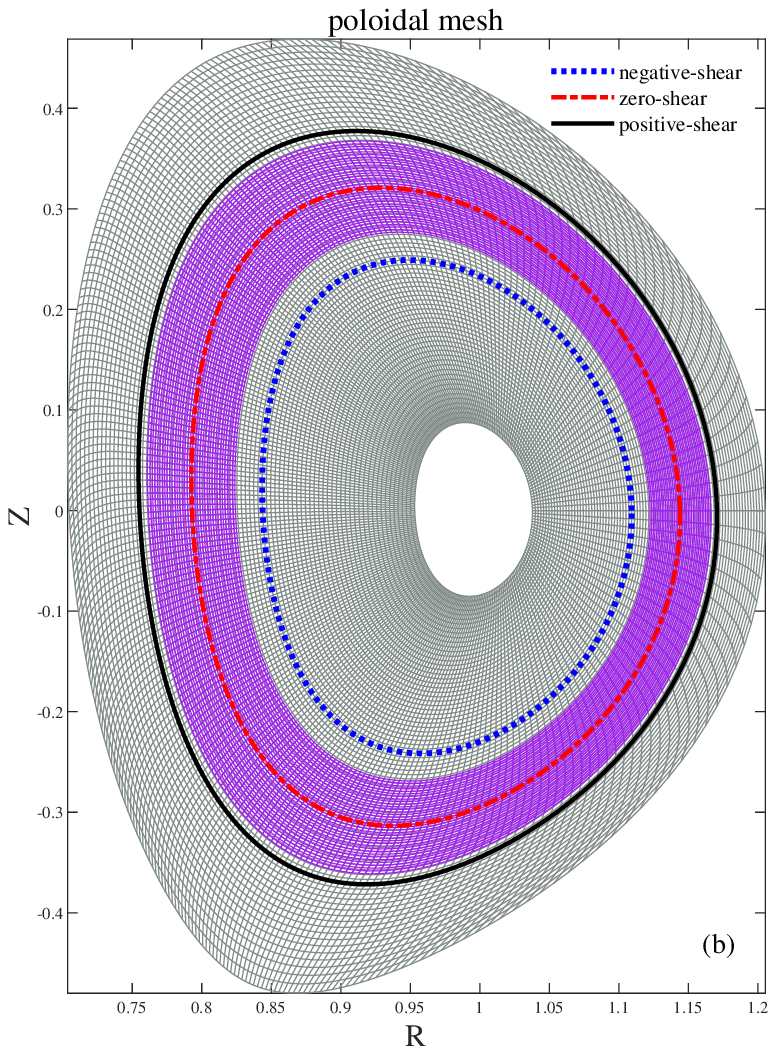}}
\caption{(a): Selection of three radial points as simulation domain centers: Point A features negative magnetic shear, Point B represents zero magnetic shear, and Point C demonstrates positive magnetic shear. (b): Central flux surfaces within simulation domains centered around these three distinct magnetic shears.}
\label{anquanyingzi}
\end{figure}

In this study, we conduct electrostatic global gyrokinetic simulations in the collisionless limit using the Gyrokinetic Toroidal Code (GTC) to investigate the influence of magnetic shear on ITG turbulence and transport, focusing on the zonal physics relevant to realistic fusion plasmas. The equilibrium magnetic field used in the simulations is obtained from the design of the China Fusion Engineering Test Reactor (CFETR)\cite{wan2017overview,zhuang2019progress}, which features strong shaping and reverse magnetic shear, as depicted in Figure \ref{anquanyingzi}. Experimental and simulation analyses have shown that reverse magnetic shear significantly affects particle and heat transport in the plasma through the redistribution of mode rational surfaces and the suppression of linear instabilities \cite{deng2009properties}. 

This study further reveals that negative magnetic shear leads to weaker zonal flows, which in turn diminishes the linear suppression effect and results in the vanishing of the Dimits shift. This observation highlights the crucial role of magnetic shear in regulating turbulence and transport in fusion plasmas.

To simulate strongly shaped plasmas, we implement the modified four-point average method within the GTC code\cite{duan2022gyro}. For our simulations, we select three reference magnetic surfaces at the centers of the simulation domains, denoted as A, B, and C in Figure \ref{anquanyingzi}(a), representing negative, zero, and positive magnetic shear, respectively.

The simulations employ the following major parameters: magnetic field at the magnetic axis $B_{0}=6.50T$ , major radius $R_{0}=6.60m$, inverse aspect ratio $a/R_{0}=0.22$, on-axis electron density $n_0=7.80 \times 10^{19}/m^3$,  $n_{i0}=n_{0}$ due to quasi-neutrality. Deuterium ions are used with a mass of $m_D=2m_H=3674m_e$. The on-axis temperature for both electrons and ions is $T_{e0} = T_{i0} = 13.00 keV$, and the normalized minor radius $a$ is approximately equal to $569 \rho_{i}$. Here, the ion gyroradius $\rho_{i}$ is calculated as $\rho_{i}=C_{s}/\Omega_{i}$, where $C_s=\sqrt{T_{e0}/m_{D}}$ and $\Omega_{i}$ represents the deuterium cyclotron frequency.

Figure \ref{anquanyingzi}(b) illustrates the poloidal cross-section of a typical CFETR equilibrium B field, depicting a strongly shaped plasma with elongation $\kappa=2.1$ and triangularity $\delta=0.1$ at $r=0.5a$. The central flux surfaces for different simulation domains are demonstrated by blue dashed lines for negative magnetic shear, red dash-dotted lines for zero magnetic shear, and black solid lines for positive magnetic shear. Each simulation domain spans 20\% of the minor radius $a$, with $\Delta r\approx100\rho_i$. The poloidal simulation domain for zero magnetic shear is denoted by the purple shaded area. We set the simulation grids with ${\Delta}r = 0.81\rho_i$ and $r{\Delta}\theta = 0.85\rho_{i}$, and there are 20 markers per cell for the ions. The simulation time step size is ${\Delta}t = 0.10R_0/C_s$, where $C_s/R_0 = 1.19\times10^{5}s^{-1}$.

\begin{figure}
     \centering
      \subfigure{
            \includegraphics[width=7cm,height=5.1cm]{
    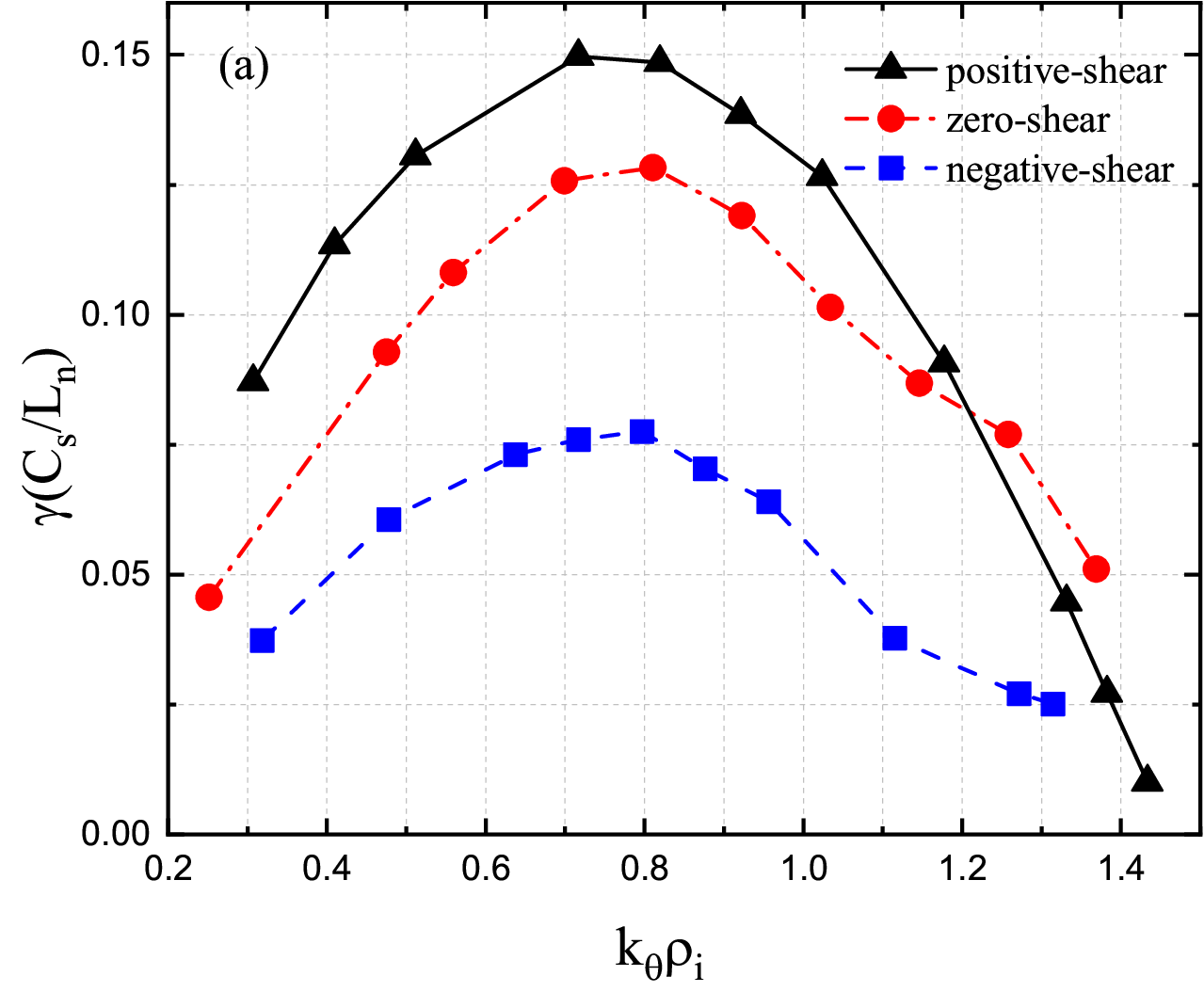}}
 \hspace{0.1in}
      \subfigure{
            \includegraphics[width=7cm,height=5cm]{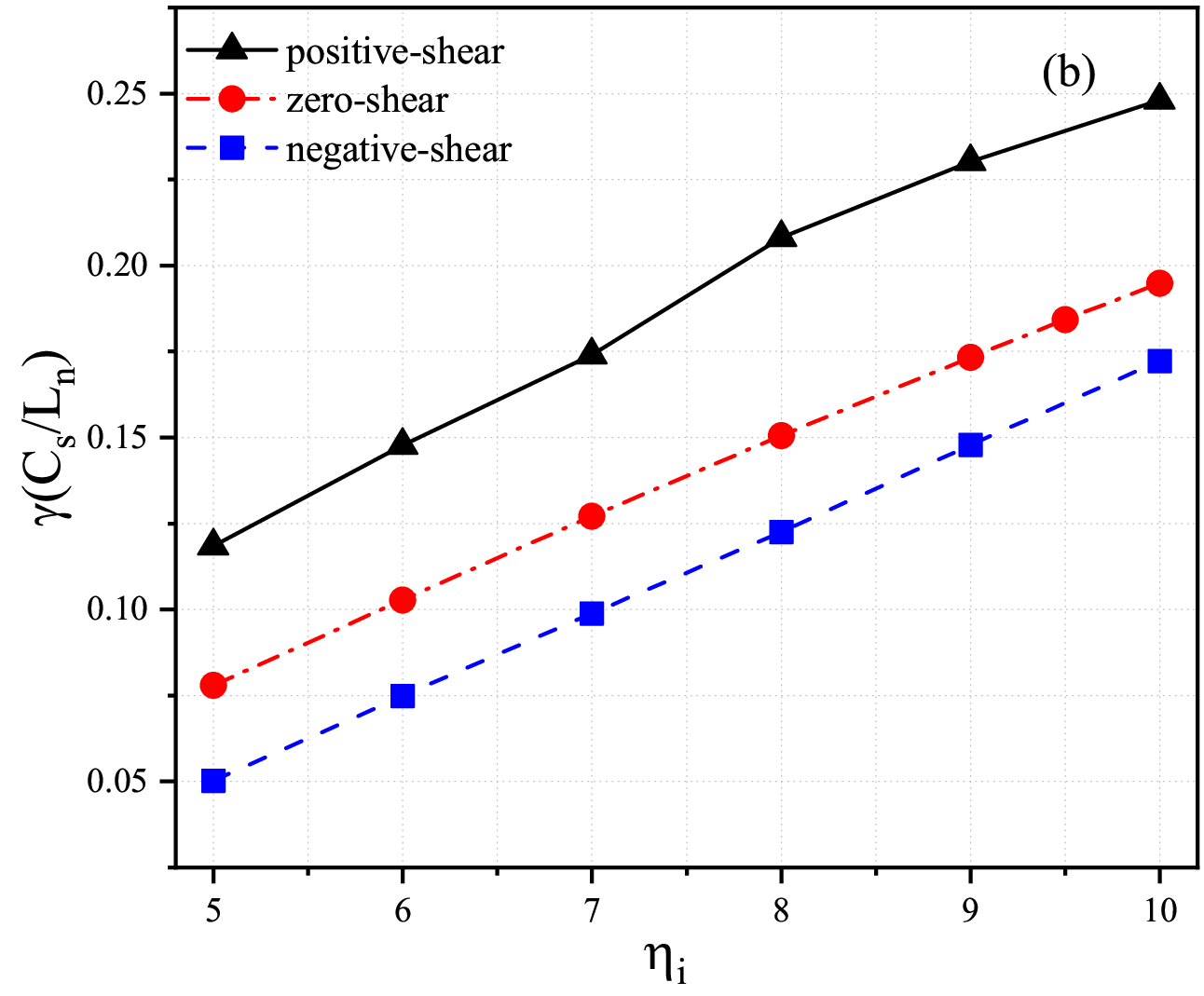}}
 \caption{(a): Variation of linear growth rate as a function of wavelength $k_{\theta}\rho_i$ (a) and ion temperature gradient or $\eta_i$ (b) for the ITG instability (adiabatic electrons) under different magnetic shear conditions.}
\label{wiuhdiaii}
\end{figure}
Here we assume adiabatic electrons in the simulations to elucidate the essential physics since grand computational challenges are posed by the large size of the CFETR tokamak.  We focus on this strongly shaped plasma and begin by investigating the linear ITG dispersion relation while varying the poloidal wavelength $k_{\theta}\rho_i$, and comparing the resulting linear growth rates and real frequencies for three different magnetic shears, as shown in Figure 1. The strong plasma shaping featured by the CFETR design, is found to significantly enhance the threshold of the ITG instability, consistent with previous simulations \cite{qi2016gyrokinetic,angelino2009role}. To excite the ITG instability, we set the temperature gradients much larger than those in the Cyclone Base (CBC) Case, with $R/L_{Ti}=R/L_{Te}=13.32$ and $R/L_{n}=2.22$ for the present study. Figure \ref{wiuhdiaii}(a) presents the ITG growth rate as a function of $k_{\theta}\rho_i$ for negative magnetic shear (blue square line), zero magnetic shear (red circle line), and positive magnetic shear (black triangle line) defined at reference points A, B, and C in Figure \ref{anquanyingzi}, respectively. The linear growth rate significantly decreases as the magnetic shear changes from positive to negative, indicating the stronger stabilizing effect of the negative shear compared to the positive shear. This result is consistent with previous gyrokinetic simulations using CBC parameters with circular magnetic flux surfaces \cite{deng2009properties}. The maximum growth rate for positive magnetic shear is $\gamma=0.15C_{s}/L_{n}$. The wavelength that maximizes the linear growth rate $k_{\theta}^{max}$ is around $k_{\theta}^{max}\rho_i=0.78$, determined by the effective perpendicular wavelength $k_{\theta}^{eff}\rho_i$, since the effective poloidal wavelength $k_{\theta}^{eff}\rho_i=\frac{k_{\theta}\rho_i}{\kappa}=0.37$ \cite{qi2016gyrokinetic}, which is consistent with the CBC case \cite{rewoldt2007linear}.

Subsequently, we examine the dependence of the maximum linear growth rate $\gamma_{max}$ on the ion temperature gradient or $\eta_i \equiv \frac{d\ln T_i}{d\ln n_i}$ for this ITG instability, presented in Figure \ref{wiuhdiaii}(b) for $R/L_{n}=2.22$ and $k_{\theta}\rho_i=0.78$. We observe that $\gamma_{max}$ increases almost linearly with $\eta_i$, with the same slope for different magnetic shears, suggesting that magnetic shear is decoupled from $\eta_i$ in determining $\gamma_{max}$. Additionally, the linear simulation results suggest that the real frequency $\omega_r \propto \omega_{*i}(1+\eta_i)$, where $\omega_{*i}$ is the ion diamagnetic frequency, consistent with theoretical predictions. These linear simulations confirm the reliability of the GTC code in simulating the ITG mode for strongly shaped, realistic tokamak plasmas.

%
\begin{figure*}[t]
   \centering
   \begin{minipage}[b]{0.3\textwidth}
     \centering
     \includegraphics[width=\textwidth]{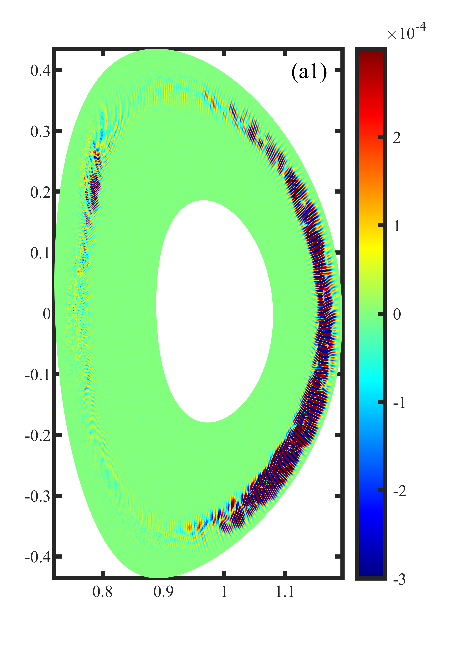}
\vspace{-5.0em}
   \end{minipage}
   \hfill
   \begin{minipage}[b]{0.3\textwidth}
     \centering
     \includegraphics[width=\textwidth]{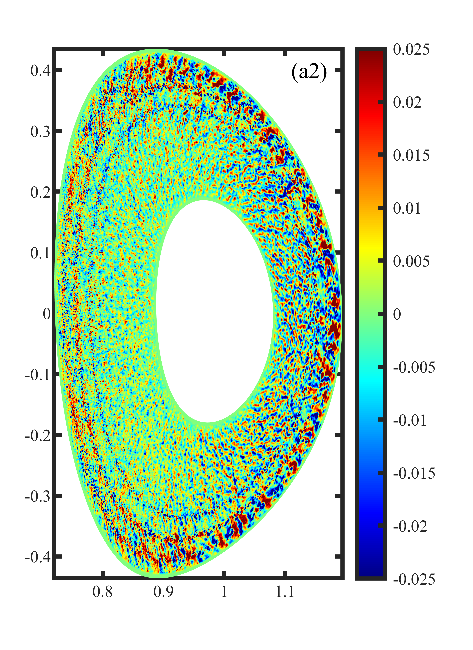}
\vspace{-5.0em}
   \end{minipage}
   \hfill
   \begin{minipage}[b]{0.3\textwidth}
     \centering
     \includegraphics[width=\textwidth]{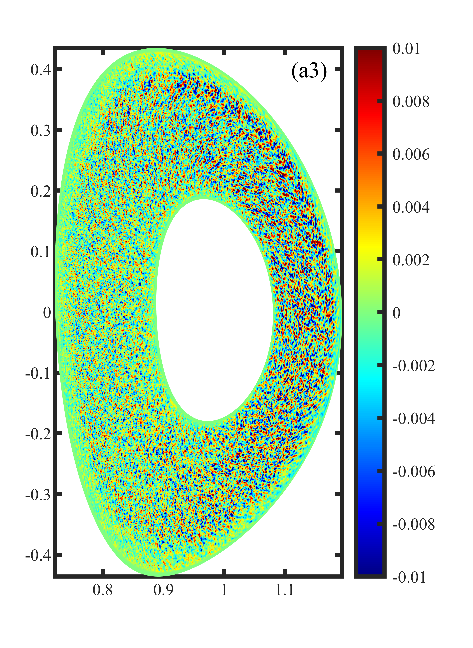}
\vspace{-5.0em}
   \end{minipage}
\vfill
   \begin{minipage}[t]{0.3\textwidth}
    \vspace{0pt}
     \centering
     \includegraphics[width=\textwidth]{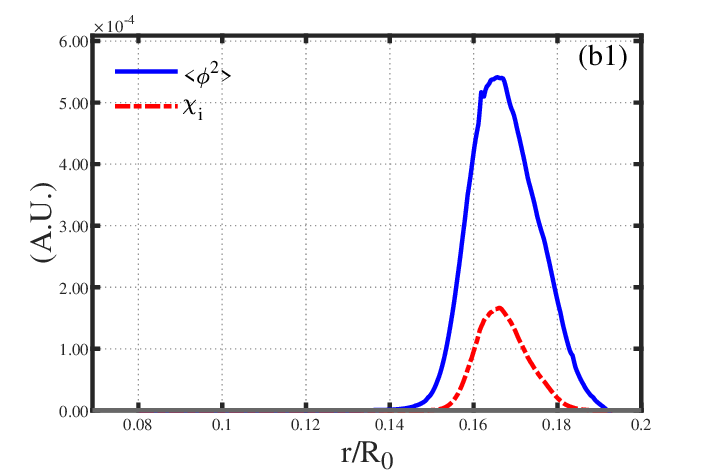}
   \end{minipage}
   \hfill
   \begin{minipage}[t]{0.3\textwidth}
     \vspace{0pt}
     \centering
     \includegraphics[width=1.15\textwidth]{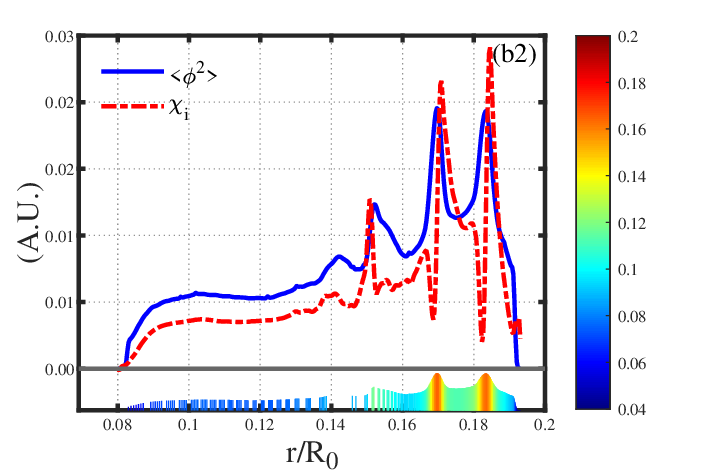}
   \end{minipage}
   \hfill
   \begin{minipage}[t]{0.3\textwidth}
     \vspace{0pt}
      \centering
     \includegraphics[width=\textwidth]{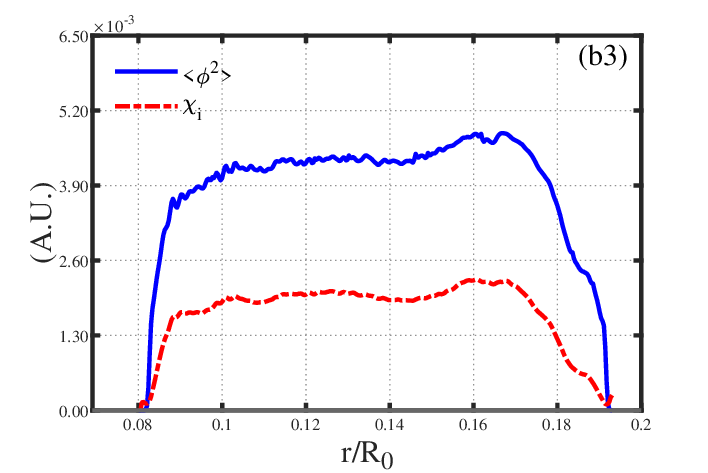}
   \end{minipage}
    \caption{(a1-a3) - poloidal turbulence contour of lTG mode under reverse magnetic shear: (a1) - linear Stage, (a2) - nonlinear stage without zonal Flow, (a3) - nonlinear stage with zonal flow. (b1-b3) - radial profiles of ion heat conductivity (red dashed line) and turbulence intensity (blue solid line) after flux surface and time averaging: (b1) - linear Stage, (b2) - nonlinear stage without zonal flow, (b3) - nonlinear stage with zonal flow.}
    \label{jixiangjiemaodkyang}
\end{figure*}
We proceed by expanding the simulation domain to encompass the entire reverse magnetic shear profile, conducting nonlinear ITG simulations within this extended radial domain. Figure \ref{jixiangjiemaodkyang} displays the 2D poloidal electrostatic potential for reverse magnetic shear, with the self-consistent zonal flow retained in the right column (a3) and artificially removed zonal flow in the middle column (a2). The linear mode structure is shown in the left column (a1). The second row presents the flux-surface-averaged turbulence intensity and ion heat conductivity for each case, indicating that ion heat transport is driven by local turbulent intensity. In the linear stage, the stronger stabilizing effect of negative shear on the linear instability confines the linear mode structure to the positive shear region. However, during the nonlinear stage, potential fluctuations spread from the positive magnetic shear region to the negative magnetic shear region, as does ion heat transport. In the absence of self-consistent zonal flow, the radial turbulence structure exhibits bursty peaks corresponding to soliton structures previously observed in research. Plotting the mode rational surface density weighted by the growth rate (Fig.3(b2)), we find that these turbulent soliton positions align with local maxima of the mode rational density profile, which implies that the turbulence soliton arises from the overlap of mode rational density and reinforcement of multi toroidal modes. In contrast, when zonal flow is self-consistently excited, the radial distribution of both turbulence intensity and ion heat transport becomes more uniform radially. This suggests that zonal flow can absorb strong local turbulence energy from densely packed mode rational surfaces, effectively eliminating the soliton structures in the ITG turbulence. Another possibility is that zonal flow propagation can expediate turbulence spreading and radial redistribution of turbulent energy.
\begin{figure}[htbp]
\includegraphics[width=8cm]{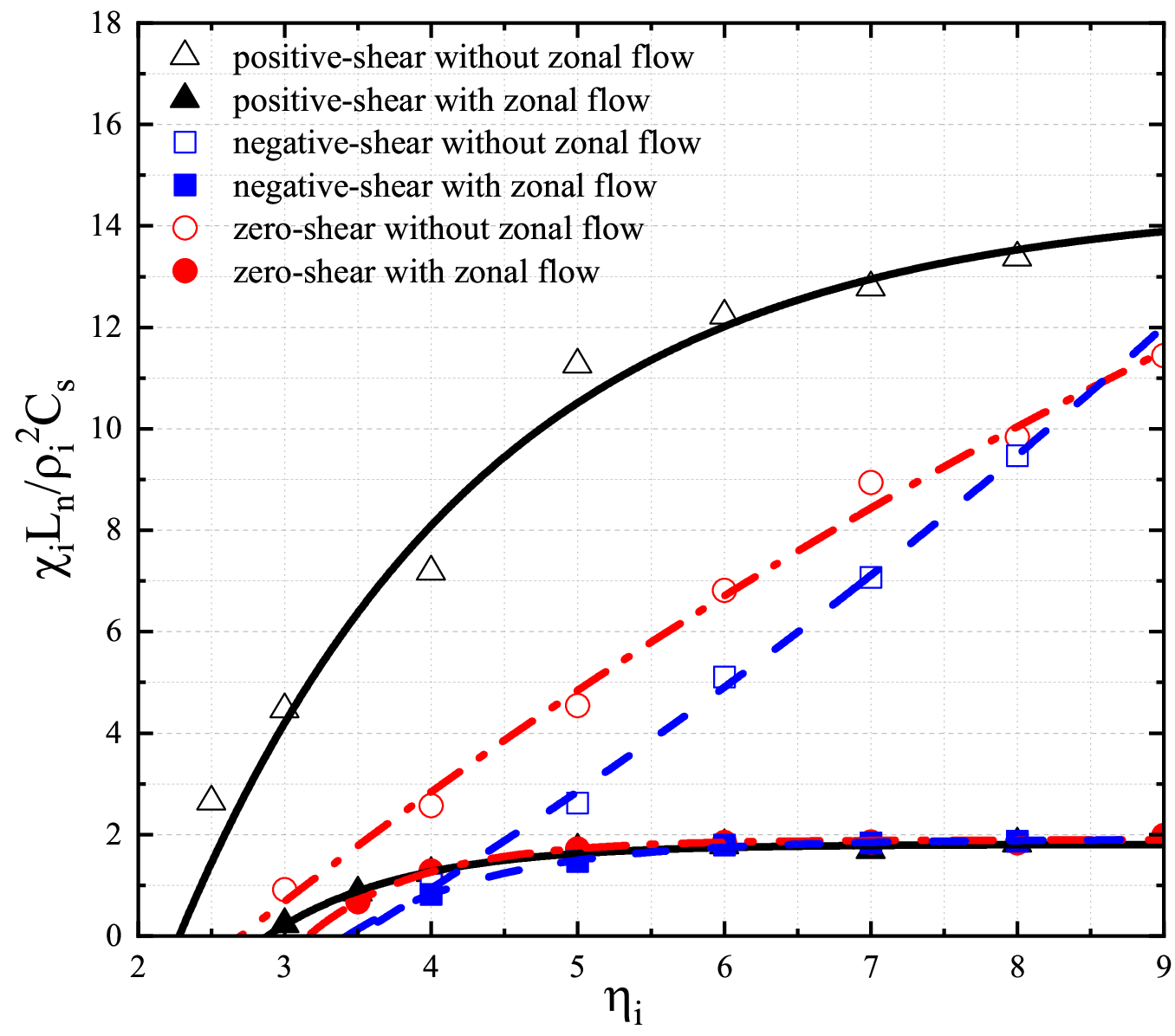}
\caption{\label{positrons} Ion thermal conductivity, $\chi_i$, varies with $\eta_i$ for various magnetic shears with or without zonal flow. }
\label{fig:9}
\end{figure}

\begin{table}[b]
\caption{\label{tab:table1}
Critical $\eta_i$ for ITG turbulence caused  by the presence or absence of zonal flows for various magnetic shears
}
\begin{ruledtabular}
\begin{tabular}{lcdr}
\textrm{shear}&
\textrm{negative}&
\multicolumn{1}{c}{\textrm{zero}}&
\textrm{positive}\\
\colrule
w/0 zonal  &3.43 &2.70 &2.28\\
with zonal  &3.45 &3.16 &2.88\\
NL upshift &0.02 &0.46 &0.60\\
\end{tabular}
\end{ruledtabular}
\end{table}

\begin{figure*}
\href{http://prst-per.aps.org/multimedia/PRSTPER/v4/i1/e010101/e010101_vid1a.mpg}{\includegraphics[width=8cm,height=6cm]{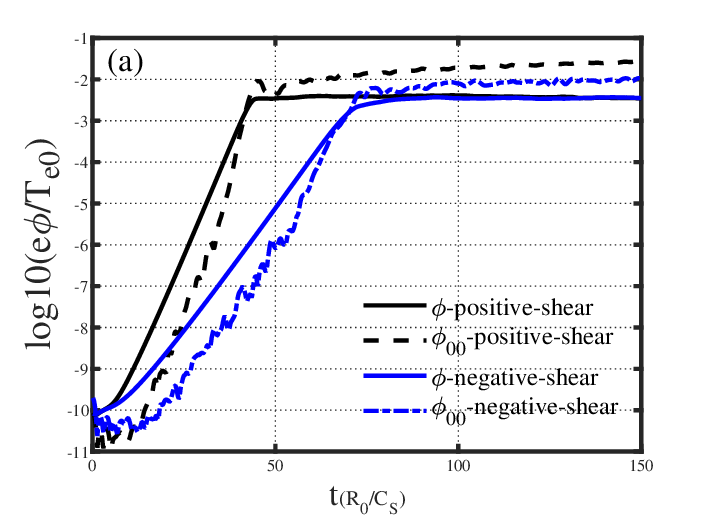}}%
 \quad
\href{http://prst-per.aps.org/multimedia/PRSTPER/v4/i1/e010101/e010101_vid1b.mpg}{\includegraphics[width=8cm,height=6.5cm]{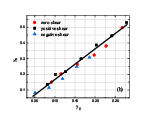}}
 \quad
\href{http://prst-per.aps.org/multimedia/PRSTPER/v4/i1/e010101/e010101_vid1b.mpg}{\includegraphics[width=8cm,height=6cm]{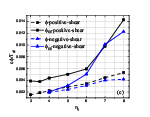}}
\setfloatlink{http://link.aps.org/multimedia/PRSTPER/v4/i1/e010101}%
 \caption{(a) - Temporal development of ITG mode and zonal flow with varied magnetic shears  for $\eta_{i}=6.0$ . (b) - Relationship between secondary growth rate ($\gamma_{s}$) or zonal flow growth rate and primary growth rate ($\gamma_{0}$)  or ITG growth rate. (c) - Saturation levels of zonal flow and ITG turbulence across different $\eta_{i}$ values.}%
\label{vid:PRSTPER.4.010101}   
\end{figure*}

Next, we begin by performing a time-averaging of the ion thermal conductivity during the nonlinear saturation stage for different magnetic shears within the shaded regions illustrated in Figure \ref{anquanyingzi}(b). The resulting saturated $\chi_i$ is then plotted as a function of the ion temperature gradient, $\eta_i$, while keeping the density gradient fixed, as shown in Figure \ref{fig:9}. In this plot, the shaped markers represent data from the GTC simulation, while the continuous lines are obtained through numerical curve fitting to illustrate the overall trend.

It is evident that, in the absence of zonal flow, the ion thermal conductivity $\chi_i$ increases rapidly with $\eta_i$. However, when self-consistent zonal flow is generated, $\chi_i$ increases at a much slower rate with $\eta_i$ and eventually reaches a plateau value, regardless of the magnetic shear. This indicates that the zonal flow serves as the dominant saturation mechanism, significantly reducing $\chi_i$, as demonstrated in Figure \ref{fig:9}. Nevertheless, this regulatory effect of the zonal flow weakens for the negative magnetic shear case near the marginal stability point.

A study reported previously that for the reverse shear case, turbulent transport does not decrease compared to the positive shear case due to turbulence spreading from positive shear regions to negative regions \cite{deng2009properties}. However, when away from marginal stability, the turbulent transport for the negative magnetic shear case resembles that of the positive magnetic shear case, which cannot be easily explained by the turbulence spreading picture \cite{deng2009properties}. This suggests that the turbulent transport reaches some fixed point, as depicted in the predator-prey turbulence-zonal model \cite{diamond2005topical}, necessitating further theoretical investigation.

The phenomenon of the nonlinear upshift for the critical $\eta_i$, known as the Dimits shift, is well-established and attributed to the presence of zonal flow. In Table \ref{tab:table1}, we present the Dimits shifts for three different magnetic shears. It is evident from the table that the Dimits shift is absent for negative magnetic shear, while for positive and zero magnetic shear, the Dimits shift is approximately the same.

The intriguing observation of the vanishing Dimits shift under negative magnetic shear merits further investigation, as the underlying physics mechanism responsible for the Dimits shift lies in the essential nonlinear interaction between turbulence and zonal flow. To shed more light on this fascinating aspect of physics, we closely examine the excitation of zonal flow during the initial stage of a multi-n mode simulation , as illustrated in Fig. 5(a). Initially, the zonal flow experiences exponential growth at a primary rate denoted by $\gamma_0$, which is equivalent to the growth rate of the ITG instability. Upon reaching a certain amplitude, the zonal flow enters a phase of growth at a different rate known as the secondary growth rate $\gamma_s$. Remarkably, this phenomenon is observed across all three magnetic shears and values of instability drive $\eta_i$. Significantly, as depicted in Figure 5(b), the secondary growth rate $\gamma_s$ consistently appears to be twice the magnitude of $\gamma_0$. This observation is further substantiated by single-n simulations involving the (n=140) mode, which corresponds to the most unstable mode. This finding strongly suggests that the force-driven process, specifically the self-interaction of single-n drift waves, acts as the dominant mechanism to excite the zonal flow. Furthermore, this force-driven mechanism remains valid in the multi-n simulations, suggesting that the realistic generation of zonal flow by turbulence is predominantly influenced by the force-driven instability, rather than the modulational instability conventionally proposed, where the growth rate of the zonal flow critically depends on the turbulence magnitude. Therefore, the existing paradigm of zonal flow-turbulence interaction demands careful reconsideration and further theoretical investigation.

In the multi-mode nonlinear simulation, the saturated zonal flow and turbulence potential are represented as a function of the instability drive $\eta_i$, as illustrated in Fig. 5(c). As the ITG mode approaches marginal stability, the magnitude of the drift wave diminishes, reaching a transport-insignificant level regardless of the magnetic shear. For the negative magnetic shear case, the zonal flow magnitude also approaches a level similar to that of the drift wave. However, in the positive magnetic shear case, the zonal flow magnitude reaches a much higher level. It is the disparity between the drift wave and the zonal flow at the point of marginal stability that governs the significance of the Dimits shift. This observation may hold the key to gaining crucial insights into revisiting the theory of zonal flow-turbulence interaction.

Yong Xiao, one of the authors, acknowledges valuable discussions and insightful suggestions from Professors Liu Chen and Fulvio Zonca, along with Dr. Zhixing Lu. Additionally, acknowledgment is extended to Dr. Qilong Ren for the provision of CFETR equilibrium data. This work is supported by by NSFC under Grant No. 11975201, and National MCF Energy R $\&$ D Program of China under Grant No. 2019YFE03060000. The computing resources have been provided by National Tianjin Supercomputing Center.

\bibliography{apssampdkyang}

\providecommand{\noopsort}[1]{}\providecommand{\singleletter}[1]{#1}%
\begin{thebibliography}{24}%
\makeatletter
\providecommand \@ifxundefined [1]{%
 \@ifx{#1\undefined}
}%
\providecommand \@ifnum [1]{%
 \ifnum #1\expandafter \@firstoftwo
 \else \expandafter \@secondoftwo
 \fi
}%
\providecommand \@ifx [1]{%
 \ifx #1\expandafter \@firstoftwo
 \else \expandafter \@secondoftwo
 \fi
}%
\providecommand \natexlab [1]{#1}%
\providecommand \enquote  [1]{``#1''}%
\providecommand \bibnamefont  [1]{#1}%
\providecommand \bibfnamefont [1]{#1}%
\providecommand \citenamefont [1]{#1}%
\providecommand \href@noop [0]{\@secondoftwo}%
\providecommand \href [0]{\begingroup \@sanitize@url \@href}%
\providecommand \@href[1]{\@@startlink{#1}\@@href}%
\providecommand \@@href[1]{\endgroup#1\@@endlink}%
\providecommand \@sanitize@url [0]{\catcode `\\12\catcode `\$12\catcode
  `\&12\catcode `\#12\catcode `\^12\catcode `\_12\catcode `\%12\relax}%
\providecommand \@@startlink[1]{}%
\providecommand \@@endlink[0]{}%
\providecommand \url  [0]{\begingroup\@sanitize@url \@url }%
\providecommand \@url [1]{\endgroup\@href {#1}{\urlprefix }}%
\providecommand \urlprefix  [0]{URL }%
\providecommand \Eprint [0]{\href }%
\providecommand \doibase [0]{https://doi.org/}%
\providecommand \selectlanguage [0]{\@gobble}%
\providecommand \bibinfo  [0]{\@secondoftwo}%
\providecommand \bibfield  [0]{\@secondoftwo}%
\providecommand \translation [1]{[#1]}%
\providecommand \BibitemOpen [0]{}%
\providecommand \bibitemStop [0]{}%
\providecommand \bibitemNoStop [0]{.\EOS\space}%
\providecommand \EOS [0]{\spacefactor3000\relax}%
\providecommand \BibitemShut  [1]{\csname bibitem#1\endcsname}%
\let\auto@bib@innerbib\@empty
\bibitem [{\citenamefont {Horton}(1999)}]{horton1999drift}%
  \BibitemOpen
  \bibfield  {author} {\bibinfo {author} {\bibfnamefont {W.}~\bibnamefont
  {Horton}},\ }\bibfield  {title} {\bibinfo {title} {Drift waves and
  transport},\ }\href@noop {} {\bibfield  {journal} {\bibinfo  {journal}
  {Reviews of Modern Physics}\ }\textbf {\bibinfo {volume} {71}},\ \bibinfo
  {pages} {735} (\bibinfo {year} {1999})}\BibitemShut {NoStop}%
\bibitem [{\citenamefont {Coppi}\ \emph {et~al.}(1967)\citenamefont {Coppi},
  \citenamefont {Rosenbluth},\ and\ \citenamefont
  {Sagdeev}}]{coppi1967instabilities}%
  \BibitemOpen
  \bibfield  {author} {\bibinfo {author} {\bibfnamefont {B.}~\bibnamefont
  {Coppi}}, \bibinfo {author} {\bibfnamefont {M.}~\bibnamefont {Rosenbluth}},\
  and\ \bibinfo {author} {\bibfnamefont {R.}~\bibnamefont {Sagdeev}},\
  }\bibfield  {title} {\bibinfo {title} {Instabilities due to temperature
  gradients in complex magnetic field configurations},\ }\href@noop {}
  {\bibfield  {journal} {\bibinfo  {journal} {The Physics of Fluids}\ }\textbf
  {\bibinfo {volume} {10}},\ \bibinfo {pages} {582} (\bibinfo {year}
  {1967})}\BibitemShut {NoStop}%
\bibitem [{\citenamefont {Romanelli}(1989)}]{romanelli1989ion}%
  \BibitemOpen
  \bibfield  {author} {\bibinfo {author} {\bibfnamefont {F.}~\bibnamefont
  {Romanelli}},\ }\bibfield  {title} {\bibinfo {title} {Ion
  temperature-gradient-driven modes and anomalous ion transport in tokamaks},\
  }\href@noop {} {\bibfield  {journal} {\bibinfo  {journal} {Physics of Fluids
  B: Plasma Physics}\ }\textbf {\bibinfo {volume} {1}},\ \bibinfo {pages}
  {1018} (\bibinfo {year} {1989})}\BibitemShut {NoStop}%
\bibitem [{\citenamefont {Dong}\ \emph {et~al.}(1992)\citenamefont {Dong},
  \citenamefont {Horton},\ and\ \citenamefont {Kim}}]{dong1992toroidal}%
  \BibitemOpen
  \bibfield  {author} {\bibinfo {author} {\bibfnamefont {J.}~\bibnamefont
  {Dong}}, \bibinfo {author} {\bibfnamefont {W.}~\bibnamefont {Horton}},\ and\
  \bibinfo {author} {\bibfnamefont {J.}~\bibnamefont {Kim}},\ }\bibfield
  {title} {\bibinfo {title} {Toroidal kinetic $\eta$ i-mode study in
  high-temperature plasmas},\ }\href@noop {} {\bibfield  {journal} {\bibinfo
  {journal} {Physics of Fluids B: Plasma Physics}\ }\textbf {\bibinfo {volume}
  {4}},\ \bibinfo {pages} {1867} (\bibinfo {year} {1992})}\BibitemShut
  {NoStop}%
\bibitem [{\citenamefont {Dorland}\ and\ \citenamefont
  {Hammett}(1993)}]{dorland1993gyrofluid}%
  \BibitemOpen
  \bibfield  {author} {\bibinfo {author} {\bibfnamefont {W.}~\bibnamefont
  {Dorland}}\ and\ \bibinfo {author} {\bibfnamefont {G.}~\bibnamefont
  {Hammett}},\ }\bibfield  {title} {\bibinfo {title} {Gyrofluid turbulence
  models with kinetic effects},\ }\href@noop {} {\bibfield  {journal} {\bibinfo
   {journal} {Physics of Fluids B: Plasma Physics}\ }\textbf {\bibinfo {volume}
  {5}},\ \bibinfo {pages} {812} (\bibinfo {year} {1993})}\BibitemShut {NoStop}%
\bibitem [{\citenamefont {Doyle}\ \emph {et~al.}(2007)\citenamefont {Doyle},
  \citenamefont {Houlberg}, \citenamefont {Kamada}, \citenamefont {Mukhovatov},
  \citenamefont {Osborne}, \citenamefont {Polevoi}, \citenamefont {Bateman},
  \citenamefont {Connor}, \citenamefont {Cordey}, \citenamefont {Fujita} \emph
  {et~al.}}]{doyle2007plasma}%
  \BibitemOpen
  \bibfield  {author} {\bibinfo {author} {\bibfnamefont {E.}~\bibnamefont
  {Doyle}}, \bibinfo {author} {\bibfnamefont {W.}~\bibnamefont {Houlberg}},
  \bibinfo {author} {\bibfnamefont {Y.}~\bibnamefont {Kamada}}, \bibinfo
  {author} {\bibfnamefont {V.}~\bibnamefont {Mukhovatov}}, \bibinfo {author}
  {\bibfnamefont {T.}~\bibnamefont {Osborne}}, \bibinfo {author} {\bibfnamefont
  {A.}~\bibnamefont {Polevoi}}, \bibinfo {author} {\bibfnamefont
  {G.}~\bibnamefont {Bateman}}, \bibinfo {author} {\bibfnamefont
  {J.}~\bibnamefont {Connor}}, \bibinfo {author} {\bibfnamefont
  {J.}~\bibnamefont {Cordey}}, \bibinfo {author} {\bibfnamefont
  {T.}~\bibnamefont {Fujita}}, \emph {et~al.},\ }\bibfield  {title} {\bibinfo
  {title} {Plasma confinement and transport},\ }\href@noop {} {\bibfield
  {journal} {\bibinfo  {journal} {Nuclear Fusion}\ }\textbf {\bibinfo {volume}
  {47}},\ \bibinfo {pages} {S18} (\bibinfo {year} {2007})}\BibitemShut
  {NoStop}%
\bibitem [{\citenamefont {Lin}\ \emph {et~al.}(1998)\citenamefont {Lin},
  \citenamefont {Hahm}, \citenamefont {Lee}, \citenamefont {Tang},\ and\
  \citenamefont {White}}]{lin1998turbulent}%
  \BibitemOpen
  \bibfield  {author} {\bibinfo {author} {\bibfnamefont {Z.}~\bibnamefont
  {Lin}}, \bibinfo {author} {\bibfnamefont {T.~S.}\ \bibnamefont {Hahm}},
  \bibinfo {author} {\bibfnamefont {W.}~\bibnamefont {Lee}}, \bibinfo {author}
  {\bibfnamefont {W.~M.}\ \bibnamefont {Tang}},\ and\ \bibinfo {author}
  {\bibfnamefont {R.~B.}\ \bibnamefont {White}},\ }\bibfield  {title} {\bibinfo
  {title} {Turbulent transport reduction by zonal flows: Massively parallel
  simulations},\ }\href@noop {} {\bibfield  {journal} {\bibinfo  {journal}
  {Science}\ }\textbf {\bibinfo {volume} {281}},\ \bibinfo {pages} {1835}
  (\bibinfo {year} {1998})}\BibitemShut {NoStop}%
\bibitem [{\citenamefont {Rosenbluth}\ and\ \citenamefont
  {Hinton}(1998)}]{rosenbluth1998poloidal}%
  \BibitemOpen
  \bibfield  {author} {\bibinfo {author} {\bibfnamefont {M.}~\bibnamefont
  {Rosenbluth}}\ and\ \bibinfo {author} {\bibfnamefont {F.}~\bibnamefont
  {Hinton}},\ }\bibfield  {title} {\bibinfo {title} {Poloidal flow driven by
  ion-temperature-gradient turbulence in tokamaks},\ }\href@noop {} {\bibfield
  {journal} {\bibinfo  {journal} {Physical review letters}\ }\textbf {\bibinfo
  {volume} {80}},\ \bibinfo {pages} {724} (\bibinfo {year} {1998})}\BibitemShut
  {NoStop}%
\bibitem [{\citenamefont {Chen}\ \emph {et~al.}(2000)\citenamefont {Chen},
  \citenamefont {Lin},\ and\ \citenamefont {White}}]{chen2000excitation}%
  \BibitemOpen
  \bibfield  {author} {\bibinfo {author} {\bibfnamefont {L.}~\bibnamefont
  {Chen}}, \bibinfo {author} {\bibfnamefont {Z.}~\bibnamefont {Lin}},\ and\
  \bibinfo {author} {\bibfnamefont {R.}~\bibnamefont {White}},\ }\bibfield
  {title} {\bibinfo {title} {Excitation of zonal flow by drift waves in
  toroidal plasmas},\ }\href@noop {} {\bibfield  {journal} {\bibinfo  {journal}
  {Physics of Plasmas}\ }\textbf {\bibinfo {volume} {7}},\ \bibinfo {pages}
  {3129} (\bibinfo {year} {2000})}\BibitemShut {NoStop}%
\bibitem [{\citenamefont {Rogers}\ \emph {et~al.}(2000)\citenamefont {Rogers},
  \citenamefont {Dorland},\ and\ \citenamefont
  {Kotschenreuther}}]{rogers2000generation}%
  \BibitemOpen
  \bibfield  {author} {\bibinfo {author} {\bibfnamefont {B.}~\bibnamefont
  {Rogers}}, \bibinfo {author} {\bibfnamefont {W.}~\bibnamefont {Dorland}},\
  and\ \bibinfo {author} {\bibfnamefont {M.}~\bibnamefont {Kotschenreuther}},\
  }\bibfield  {title} {\bibinfo {title} {Generation and stability of zonal
  flows in ion-temperature-gradient mode turbulence},\ }\href@noop {}
  {\bibfield  {journal} {\bibinfo  {journal} {Physical review letters}\
  }\textbf {\bibinfo {volume} {85}},\ \bibinfo {pages} {5336} (\bibinfo {year}
  {2000})}\BibitemShut {NoStop}%
\bibitem [{\citenamefont {Dimits}\ \emph {et~al.}(1996)\citenamefont {Dimits},
  \citenamefont {Williams}, \citenamefont {Byers},\ and\ \citenamefont
  {Cohen}}]{dimits1996scalings}%
  \BibitemOpen
  \bibfield  {author} {\bibinfo {author} {\bibfnamefont {A.}~\bibnamefont
  {Dimits}}, \bibinfo {author} {\bibfnamefont {T.}~\bibnamefont {Williams}},
  \bibinfo {author} {\bibfnamefont {J.}~\bibnamefont {Byers}},\ and\ \bibinfo
  {author} {\bibfnamefont {B.}~\bibnamefont {Cohen}},\ }\bibfield  {title}
  {\bibinfo {title} {Scalings of ion-temperature-gradient-driven anomalous
  transport in tokamaks},\ }\href@noop {} {\bibfield  {journal} {\bibinfo
  {journal} {Physical review letters}\ }\textbf {\bibinfo {volume} {77}},\
  \bibinfo {pages} {71} (\bibinfo {year} {1996})}\BibitemShut {NoStop}%
\bibitem [{\citenamefont {Dimits}\ \emph {et~al.}(2000)\citenamefont {Dimits},
  \citenamefont {Bateman}, \citenamefont {Beer}, \citenamefont {Cohen},
  \citenamefont {Dorland}, \citenamefont {Hammett}, \citenamefont {Kim},
  \citenamefont {Kinsey}, \citenamefont {Kotschenreuther}, \citenamefont
  {Kritz} \emph {et~al.}}]{dimits2000comparisons}%
  \BibitemOpen
  \bibfield  {author} {\bibinfo {author} {\bibfnamefont {A.~M.}\ \bibnamefont
  {Dimits}}, \bibinfo {author} {\bibfnamefont {G.}~\bibnamefont {Bateman}},
  \bibinfo {author} {\bibfnamefont {M.}~\bibnamefont {Beer}}, \bibinfo {author}
  {\bibfnamefont {B.}~\bibnamefont {Cohen}}, \bibinfo {author} {\bibfnamefont
  {W.}~\bibnamefont {Dorland}}, \bibinfo {author} {\bibfnamefont
  {G.}~\bibnamefont {Hammett}}, \bibinfo {author} {\bibfnamefont
  {C.}~\bibnamefont {Kim}}, \bibinfo {author} {\bibfnamefont {J.}~\bibnamefont
  {Kinsey}}, \bibinfo {author} {\bibfnamefont {M.}~\bibnamefont
  {Kotschenreuther}}, \bibinfo {author} {\bibfnamefont {A.}~\bibnamefont
  {Kritz}}, \emph {et~al.},\ }\bibfield  {title} {\bibinfo {title} {Comparisons
  and physics basis of tokamak transport models and turbulence simulations},\
  }\href@noop {} {\bibfield  {journal} {\bibinfo  {journal} {Physics of
  Plasmas}\ }\textbf {\bibinfo {volume} {7}},\ \bibinfo {pages} {969} (\bibinfo
  {year} {2000})}\BibitemShut {NoStop}%
\bibitem [{\citenamefont {Diamond}\ \emph {et~al.}(2005)\citenamefont
  {Diamond}, \citenamefont {Itoh}, \citenamefont {Itoh},\ and\ \citenamefont
  {Hahm}}]{diamond2005topical}%
  \BibitemOpen
  \bibfield  {author} {\bibinfo {author} {\bibfnamefont {P.}~\bibnamefont
  {Diamond}}, \bibinfo {author} {\bibfnamefont {S.-I.}\ \bibnamefont {Itoh}},
  \bibinfo {author} {\bibfnamefont {K.}~\bibnamefont {Itoh}},\ and\ \bibinfo
  {author} {\bibfnamefont {T.}~\bibnamefont {Hahm}},\ }\bibfield  {title}
  {\bibinfo {title} {Topical review: zonal flows in plasma—a review},\
  }\href@noop {} {\bibfield  {journal} {\bibinfo  {journal} {Plasma Physics and
  Controlled Fusion}\ }\textbf {\bibinfo {volume} {47}},\ \bibinfo {pages}
  {R35} (\bibinfo {year} {2005})}\BibitemShut {NoStop}%
\bibitem [{\citenamefont {Zhu}\ \emph {et~al.}(2020)\citenamefont {Zhu},
  \citenamefont {Zhou},\ and\ \citenamefont {Dodin}}]{zhu2020theory}%
  \BibitemOpen
  \bibfield  {author} {\bibinfo {author} {\bibfnamefont {H.}~\bibnamefont
  {Zhu}}, \bibinfo {author} {\bibfnamefont {Y.}~\bibnamefont {Zhou}},\ and\
  \bibinfo {author} {\bibfnamefont {I.}~\bibnamefont {Dodin}},\ }\bibfield
  {title} {\bibinfo {title} {Theory of the tertiary instability and the dimits
  shift from reduced drift-wave models},\ }\href@noop {} {\bibfield  {journal}
  {\bibinfo  {journal} {Physical Review Letters}\ }\textbf {\bibinfo {volume}
  {124}},\ \bibinfo {pages} {055002} (\bibinfo {year} {2020})}\BibitemShut
  {NoStop}%
\bibitem [{\citenamefont {Mikkelsen}\ and\ \citenamefont
  {Dorland}(2008)}]{mikkelsen2008dimits}%
  \BibitemOpen
  \bibfield  {author} {\bibinfo {author} {\bibfnamefont {D.}~\bibnamefont
  {Mikkelsen}}\ and\ \bibinfo {author} {\bibfnamefont {W.}~\bibnamefont
  {Dorland}},\ }\bibfield  {title} {\bibinfo {title} {Dimits shift in realistic
  gyrokinetic plasma-turbulence simulations},\ }\href@noop {} {\bibfield
  {journal} {\bibinfo  {journal} {Physical review letters}\ }\textbf {\bibinfo
  {volume} {101}},\ \bibinfo {pages} {135003} (\bibinfo {year}
  {2008})}\BibitemShut {NoStop}%
\bibitem [{\citenamefont {Hinton}\ and\ \citenamefont
  {Rosenbluth}(1999)}]{hinton1999dynamics}%
  \BibitemOpen
  \bibfield  {author} {\bibinfo {author} {\bibfnamefont {F.}~\bibnamefont
  {Hinton}}\ and\ \bibinfo {author} {\bibfnamefont {M.}~\bibnamefont
  {Rosenbluth}},\ }\bibfield  {title} {\bibinfo {title} {Dynamics of
  axisymmetric and poloidal flows in tokamaks},\ }\href@noop {} {\bibfield
  {journal} {\bibinfo  {journal} {Plasma physics and controlled fusion}\
  }\textbf {\bibinfo {volume} {41}},\ \bibinfo {pages} {A653} (\bibinfo {year}
  {1999})}\BibitemShut {NoStop}%
\bibitem [{\citenamefont {Xiao}\ \emph {et~al.}(2007)\citenamefont {Xiao},
  \citenamefont {Catto},\ and\ \citenamefont {Dorland}}]{xiao2007effects}%
  \BibitemOpen
  \bibfield  {author} {\bibinfo {author} {\bibfnamefont {Y.}~\bibnamefont
  {Xiao}}, \bibinfo {author} {\bibfnamefont {P.~J.}\ \bibnamefont {Catto}},\
  and\ \bibinfo {author} {\bibfnamefont {W.}~\bibnamefont {Dorland}},\
  }\bibfield  {title} {\bibinfo {title} {Effects of finite poloidal gyroradius,
  shaping, and collisions on the zonal flow residual},\ }\href@noop {}
  {\bibfield  {journal} {\bibinfo  {journal} {Physics of plasmas}\ }\textbf
  {\bibinfo {volume} {14}},\ \bibinfo {pages} {055910} (\bibinfo {year}
  {2007})}\BibitemShut {NoStop}%
\bibitem [{\citenamefont {Wan}\ \emph {et~al.}(2017)\citenamefont {Wan},
  \citenamefont {Li}, \citenamefont {Liu}, \citenamefont {Wang}, \citenamefont
  {Chan}, \citenamefont {Chen}, \citenamefont {Duan}, \citenamefont {Fu},
  \citenamefont {Gao}, \citenamefont {Feng} \emph {et~al.}}]{wan2017overview}%
  \BibitemOpen
  \bibfield  {author} {\bibinfo {author} {\bibfnamefont {Y.}~\bibnamefont
  {Wan}}, \bibinfo {author} {\bibfnamefont {J.}~\bibnamefont {Li}}, \bibinfo
  {author} {\bibfnamefont {Y.}~\bibnamefont {Liu}}, \bibinfo {author}
  {\bibfnamefont {X.}~\bibnamefont {Wang}}, \bibinfo {author} {\bibfnamefont
  {V.}~\bibnamefont {Chan}}, \bibinfo {author} {\bibfnamefont {C.}~\bibnamefont
  {Chen}}, \bibinfo {author} {\bibfnamefont {X.}~\bibnamefont {Duan}}, \bibinfo
  {author} {\bibfnamefont {P.}~\bibnamefont {Fu}}, \bibinfo {author}
  {\bibfnamefont {X.}~\bibnamefont {Gao}}, \bibinfo {author} {\bibfnamefont
  {K.}~\bibnamefont {Feng}}, \emph {et~al.},\ }\bibfield  {title} {\bibinfo
  {title} {Overview of the present progress and activities on the cfetr},\
  }\href@noop {} {\bibfield  {journal} {\bibinfo  {journal} {Nuclear Fusion}\
  }\textbf {\bibinfo {volume} {57}},\ \bibinfo {pages} {102009} (\bibinfo
  {year} {2017})}\BibitemShut {NoStop}%
\bibitem [{\citenamefont {Zhuang}\ \emph {et~al.}(2019)\citenamefont {Zhuang},
  \citenamefont {Li}, \citenamefont {Li}, \citenamefont {Wan}, \citenamefont
  {Liu}, \citenamefont {Wang}, \citenamefont {Song}, \citenamefont {Chan},
  \citenamefont {Yang}, \citenamefont {Wan} \emph
  {et~al.}}]{zhuang2019progress}%
  \BibitemOpen
  \bibfield  {author} {\bibinfo {author} {\bibfnamefont {G.}~\bibnamefont
  {Zhuang}}, \bibinfo {author} {\bibfnamefont {G.}~\bibnamefont {Li}}, \bibinfo
  {author} {\bibfnamefont {J.}~\bibnamefont {Li}}, \bibinfo {author}
  {\bibfnamefont {Y.}~\bibnamefont {Wan}}, \bibinfo {author} {\bibfnamefont
  {Y.}~\bibnamefont {Liu}}, \bibinfo {author} {\bibfnamefont {X.}~\bibnamefont
  {Wang}}, \bibinfo {author} {\bibfnamefont {Y.}~\bibnamefont {Song}}, \bibinfo
  {author} {\bibfnamefont {V.}~\bibnamefont {Chan}}, \bibinfo {author}
  {\bibfnamefont {Q.}~\bibnamefont {Yang}}, \bibinfo {author} {\bibfnamefont
  {B.}~\bibnamefont {Wan}}, \emph {et~al.},\ }\bibfield  {title} {\bibinfo
  {title} {Progress of the cfetr design},\ }\href@noop {} {\bibfield  {journal}
  {\bibinfo  {journal} {Nuclear Fusion}\ }\textbf {\bibinfo {volume} {59}},\
  \bibinfo {pages} {112010} (\bibinfo {year} {2019})}\BibitemShut {NoStop}%
\bibitem [{\citenamefont {Deng}\ and\ \citenamefont
  {Lin}(2009)}]{deng2009properties}%
  \BibitemOpen
  \bibfield  {author} {\bibinfo {author} {\bibfnamefont {W.}~\bibnamefont
  {Deng}}\ and\ \bibinfo {author} {\bibfnamefont {Z.}~\bibnamefont {Lin}},\
  }\bibfield  {title} {\bibinfo {title} {Properties of microturbulence in
  toroidal plasmas with reversed magnetic shear},\ }\href@noop {} {\bibfield
  {journal} {\bibinfo  {journal} {Physics of Plasmas}\ }\textbf {\bibinfo
  {volume} {16}} (\bibinfo {year} {2009})}\BibitemShut {NoStop}%
\bibitem [{\citenamefont {Duan}\ \emph {et~al.}(2022)\citenamefont {Duan},
  \citenamefont {Xiao},\ and\ \citenamefont {Lin}}]{duan2022gyro}%
  \BibitemOpen
  \bibfield  {author} {\bibinfo {author} {\bibfnamefont {Y.}~\bibnamefont
  {Duan}}, \bibinfo {author} {\bibfnamefont {Y.}~\bibnamefont {Xiao}},\ and\
  \bibinfo {author} {\bibfnamefont {Z.}~\bibnamefont {Lin}},\ }\bibfield
  {title} {\bibinfo {title} {Gyro-average method for global gyrokinetic
  particle simulation in realistic tokamak geometry},\ }\href@noop {}
  {\bibfield  {journal} {\bibinfo  {journal} {Plasma Physics and Controlled
  Fusion}\ }\textbf {\bibinfo {volume} {64}},\ \bibinfo {pages} {045018}
  (\bibinfo {year} {2022})}\BibitemShut {NoStop}%
\bibitem [{\citenamefont {Qi}\ \emph {et~al.}(2016)\citenamefont {Qi},
  \citenamefont {Kwon}, \citenamefont {Hahm},\ and\ \citenamefont
  {Jo}}]{qi2016gyrokinetic}%
  \BibitemOpen
  \bibfield  {author} {\bibinfo {author} {\bibfnamefont {L.}~\bibnamefont
  {Qi}}, \bibinfo {author} {\bibfnamefont {J.}~\bibnamefont {Kwon}}, \bibinfo
  {author} {\bibfnamefont {T.}~\bibnamefont {Hahm}},\ and\ \bibinfo {author}
  {\bibfnamefont {G.}~\bibnamefont {Jo}},\ }\bibfield  {title} {\bibinfo
  {title} {Gyrokinetic simulations of electrostatic microinstabilities with
  bounce-averaged kinetic electrons for shaped tokamak plasmas},\ }\href@noop
  {} {\bibfield  {journal} {\bibinfo  {journal} {Physics of Plasmas}\ }\textbf
  {\bibinfo {volume} {23}},\ \bibinfo {pages} {062513} (\bibinfo {year}
  {2016})}\BibitemShut {NoStop}%
\bibitem [{\citenamefont {Angelino}\ \emph {et~al.}(2009)\citenamefont
  {Angelino}, \citenamefont {Garbet}, \citenamefont {Villard}, \citenamefont
  {Bottino}, \citenamefont {Jolliet}, \citenamefont {Ghendrih}, \citenamefont
  {Grandgirard}, \citenamefont {McMillan}, \citenamefont {Sarazin},
  \citenamefont {Dif-Pradalier} \emph {et~al.}}]{angelino2009role}%
  \BibitemOpen
  \bibfield  {author} {\bibinfo {author} {\bibfnamefont {P.}~\bibnamefont
  {Angelino}}, \bibinfo {author} {\bibfnamefont {X.}~\bibnamefont {Garbet}},
  \bibinfo {author} {\bibfnamefont {L.}~\bibnamefont {Villard}}, \bibinfo
  {author} {\bibfnamefont {A.}~\bibnamefont {Bottino}}, \bibinfo {author}
  {\bibfnamefont {S.}~\bibnamefont {Jolliet}}, \bibinfo {author} {\bibfnamefont
  {P.}~\bibnamefont {Ghendrih}}, \bibinfo {author} {\bibfnamefont
  {V.}~\bibnamefont {Grandgirard}}, \bibinfo {author} {\bibfnamefont {B.~F.}\
  \bibnamefont {McMillan}}, \bibinfo {author} {\bibfnamefont {Y.}~\bibnamefont
  {Sarazin}}, \bibinfo {author} {\bibfnamefont {G.}~\bibnamefont
  {Dif-Pradalier}}, \emph {et~al.},\ }\bibfield  {title} {\bibinfo {title}
  {Role of plasma elongation on turbulent transport in magnetically confined
  plasmas},\ }\href@noop {} {\bibfield  {journal} {\bibinfo  {journal}
  {Physical review letters}\ }\textbf {\bibinfo {volume} {102}},\ \bibinfo
  {pages} {195002} (\bibinfo {year} {2009})}\BibitemShut {NoStop}%
\bibitem [{\citenamefont {Rewoldt}\ \emph {et~al.}(2007)\citenamefont
  {Rewoldt}, \citenamefont {Lin},\ and\ \citenamefont
  {Idomura}}]{rewoldt2007linear}%
  \BibitemOpen
  \bibfield  {author} {\bibinfo {author} {\bibfnamefont {G.}~\bibnamefont
  {Rewoldt}}, \bibinfo {author} {\bibfnamefont {Z.}~\bibnamefont {Lin}},\ and\
  \bibinfo {author} {\bibfnamefont {Y.}~\bibnamefont {Idomura}},\ }\bibfield
  {title} {\bibinfo {title} {Linear comparison of gyrokinetic codes with
  trapped electrons},\ }\href@noop {} {\bibfield  {journal} {\bibinfo
  {journal} {Computer Physics Communications}\ }\textbf {\bibinfo {volume}
  {177}},\ \bibinfo {pages} {775} (\bibinfo {year} {2007})}\BibitemShut
  {NoStop}%
\end{thebibliography}%

\end{document}